\documentclass[12pt]{article}
\usepackage{amsfonts}
\usepackage{amssymb}%,letterspace}
\usepackage{graphics,psboxit,amsmath}
\usepackage{subfigure}
\usepackage{graphicx}
\usepackage{verbatim}
\usepackage{epic}

\def\hybrid{\topmargin 0pt \oddsidemargin 0pt %%%%%%%%%%%%%% Archive-30pt
        \headheight 0pt \headsep 0pt
        \textwidth 16,5cm % A4 paper
        \textheight 23,5cm % A4 paper
        \marginparwidth .875in
        \parskip 5pt plus 1pt \jot = 1.5ex}

\hybrid

\def\baselinestretch{1.2}

\catcode`\@=11

\def\marginnote#1{}
\newcount\hour
\newcount\minute
\newtoks\amorpm
\hour=\time\divide\hour by60
\minute=\time{\multiply\hour by60 \global\advance\minute by-\hour}
\edef\standardtime{{\ifnum\hour<12 \global\amorpm={am}%
        \else\global\amorpm={pm}\advance\hour by-12 \fi
        \ifnum\hour=0 \hour=12 \fi
        \number\hour:\ifnum\minute<10 0\fi\number\minute\the\amorpm}}
\edef\militarytime{\number\hour:\ifnum\minute<10 0\fi\number\minute}
\def\draftlabel#1{{\@bsphack\if@filesw {\let\thepage\relax
   \xdef\@gtempa{\write\@auxout{\string
      \newlabel{#1}{{\@currentlabel}{\thepage}}}}}\@gtempa
   \if@nobreak \ifvmode\nobreak\fi\fi\fi\@esphack}
        \gdef\@eqnlabel{#1}}
\def\@eqnlabel{}
\def\@vacuum{}
\def\draftmarginnote#1{\marginpar{\raggedright\scriptsize\tt#1}}

\def\draft{\oddsidemargin -.5truein
        \def\@oddfoot{\sl preliminary draft \hfil
        \rm\thepage\hfil\sl\today\quad\militarytime}
        \let\@evenfoot\@oddfoot \overfullrule 3pt
        \let\label=\draftlabel
        \let\marginnote=\draftmarginnote
   \def\@eqnnum{(\theequation)\rlap{\kern\marginparsep\tt\@eqnlabel}%
\global\let\@eqnlabel\@vacuum} }

\def\draft2{
        \def\@oddfoot{\sl preliminary draft \hfil
        \rm\thepage\hfil\sl\today\quad\militarytime}
        \let\@evenfoot\@oddfoot \overfullrule 3pt
        \let\label=\draftlabel
        \let\marginnote=\draftmarginnote
   \def\@eqnnum{(\theequation)\rlap{\kern\marginparsep\tt\@eqnlabel}%
\global\let\@eqnlabel\@vacuum} }

\def\preprint{\twocolumn\sloppy\flushbottom\parindent 2em
        \leftmargini 2em\leftmarginv .5em\leftmarginvi .5em
        \oddsidemargin -.5in \evensidemargin -.5in
        \columnsep .4in \footheight 0pt
        \textwidth 10.in \topmargin -.4in
        \headheight 12pt \topskip .4in
        \textheight 6.9in \footskip 0pt
        \def\@oddhead{\thepage\hfil\addtocounter{page}{1}\thepage}
        \let\@evenhead\@oddhead \def\@oddfoot{} \def\@evenfoot{} }

\def\numberbysection{\@addtoreset{equation}{section}
        \def\theequation{\thesection.\arabic{equation}}}

\def\underline#1{\relax\ifmmode\@@underline#1\else
        $\@@underline{\hbox{#1}}$\relax\fi}

\def\titlepage{\@restonecolfalse\if@twocolumn\@restonecoltrue\onecolumn
     \else \newpage \fi \thispagestyle{empty}\c@page\z@
        \def\thefootnote{\fnsymbol{footnote}} }

\def\endtitlepage{\if@restonecol\twocolumn \else \newpage \fi
        \def\thefootnote{\arabic{footnote}}
        \setcounter{footnote}{0}} %\c@footnote\z@ }

\catcode`@=12
\relax

\def\figcap{\section*{Figure Captions\markboth
        {FIGURECAPTIONS}{FIGURECAPTIONS}}\list
        {Figure \arabic{enumi}:\hfill}{\settowidth\labelwidth{Figure
999:}
        \leftmargin\labelwidth
        \advance\leftmargin\labelsep\usecounter{enumi}}}
 \relax
\def\tablecap{\section*{Table Captions\markboth
        {TABLECAPTIONS}{TABLECAPTIONS}}\list
        {Table \arabic{enumi}:\hfill}{\settowidth\labelwidth{Table
999:}
        \leftmargin\labelwidth
        \advance\leftmargin\labelsep\usecounter{enumi}}}
 \relax
\def\reflist{\section*{References\markboth
        {REFLIST}{REFLIST}}\list
        {[\arabic{enumi}]\hfill}{\settowidth\labelwidth{[999]}
        \leftmargin\labelwidth
        \advance\leftmargin\labelsep\usecounter{enumi}}}
 \relax
\makeatletter
\newcounter{pubctr}
\def\publist{\@ifnextchar[{\@publist}{\@@publist}}
\def\@publist[#1]{\list
        {[\arabic{pubctr}]\hfill}{\settowidth\labelwidth{[999]}
        \leftmargin\labelwidth
        \advance\leftmargin\labelsep
        \@nmbrlisttrue\def\@listctr{pubctr}
        \setcounter{pubctr}{#1}\addtocounter{pubctr}{-1}}}
\def\@@publist{\list
        {[\arabic{pubctr}]\hfill}{\settowidth\labelwidth{[999]}
        \leftmargin\labelwidth
        \advance\leftmargin\labelsep
        \@nmbrlisttrue\def\@listctr{pubctr}}}
 \relax
\makeatother

\def\ba{\begin{equation}}
\def\ea{\end{equation}}
\def\d{\delta}

\def\th{\theta}

\def\l{\lambda}

\def\s{\sigma}

\def\no{\noindent}
\def\hb{\hfill\break}
\def\qq{\qquad}

\def\IR{\relax{\rm I\kern-.18em R}}

\def \ha {{1\over 2}}

\begin{document}
%\draft2

%\renewcommand{\theequation}{\arabic{equation}}
%\renewcommand{\theequation}{\thesection.\arabic{equation}}

\renewcommand{\theequation}{\thesection.\arabic{equation}}
\csname @addtoreset\endcsname{equation}{section}

\newcommand{\eqn}[1]{(\ref{#1})}
\newcommand{\be}{\begin{eqnarray}}
\newcommand{\ee}{\end{eqnarray}}
\newcommand{\non}{\nonumber}

\begin{titlepage}
\strut\hfill
\vskip 1.3cm
\begin{center}

{\large \bf Integrable quantum spin chains and their\\ classical continuous counterparts}\footnote{{\tt Proceedings contribution to the Corfu Summer Institute on Elementary Particle Physics and Gravity - Workshop on Non Commutative Field Theory and Gravity, 8-12 September 2010, Corfu, Greece.
Based on a talk presented by A. Doikou.}}

\vskip 0.5in

{\bf Jean Avan$^{a}$, Anastasia Doikou$^{b}$
 and Konstadinos Sfetsos$^{b}$}\\
\vskip 0.3in

{\footnotesize $^a$ LPTM, Universite de Cergy-Pontoise (CNRS UMR 8089),\\
F-95302 Cergy-Pontoise, France}\\

\vskip .1in

{\footnotesize
$^{b}$ Department of Engineering Sciences, University of Patras,\\
GR-26500 Patras, Greece}\\

\vskip .2in

%\vskip -.15in

{\footnotesize {\tt avan@u-cergy.fr, adoikou@upatras.gr, sfetsos@upatras.gr}}\\

\end{center}

%\vfill
\vskip .6in

\centerline{\bf Abstract}

\no
We present certain classical continuum long wave-length limits
of prototype integrable quantum spin chains, and define the
corresponding construction of classical continuum Lax operators.
We also provide two specific examples, i.e. the isotropic and
anisotropic Heisenberg models.

\no

\vfill
\no

\end{titlepage}
\vfill
%\vskip .5cm
%\noindent

%\end{titlepage}
%\vfill
\eject

%\def\baselinestretch{1.2}
%\baselineskip 10 pt
%\noindent

%\tableofcontents

\def\baselinestretch{1.2}
\baselineskip 20 pt
\no

\section{Introduction}

Locally interacting discrete integrable spin chains have been
the subject of much interest since they cropped
up in string theory in the study of the AdS/CFT correspondence \cite{miha}.
Their classical, long wavelength limit, provides a connection
to continuous $\s$-models describing
particular dynamics of the string (references on this subject can be found in e.g. \cite{fradkin, string}).

\no
Our motivation for this work is to develop a Hamiltonian approach
different in its principle from the usual Lagrangian formulation of the long wavelength limit, in
order to use in cases where the latter cannot be applied.
In our approach we start from the Hamiltonian integrability formulation (quantum R-matrix and Lax matrix)
guaranteeing a priori Liouville integrability of the classical continuous models.
This is done through a Lax matrix-classical $r$-matrix formulation, provided that some consistency checks be made.
On all known specific examples it will be checked that it yields
the same results as the Lagrangian approach.
It is indeed a key result that the Poisson structure is the same, in all cases when comparison
is available, as the canonical structure derived from the long wavelength classical Lagrangian.
This thereby validates the procedure and allows to use it in more general situations where the
Lagrangian approach may not be used, in particular as a systematic way to build more general
types of classical continuous integrable models by exploiting the richness of the algebraic
approach.

\no
This contribution is based on \cite{ads}, where the interested reader can find all the details of the construction.

\section{The general setting}

In this section we outline the general
procedure for obtaining a classical Lax formulation from the classical limits of the
$R$ and monodromy matrices.

\no
A quantum $c$-number non-dynamical $R$-matrix obeys the quantum Yang--Baxter (YB) equation \cite{baxter}
\be
R_{12}\ R_{13}\ R_{23} = R_{23}\ R_{13}\ R_{12} \ ,
 \label{YBR}
\ee
where the labels $i = 1,2,3$ may include dependence on a complex
spectral parameter $\lambda_i$. The auxiliary spaces are in this case
loop-spaces $V_i \otimes C(\lambda_i)$, where $V_i$ are (isomorphic) finite-dimensional vector spaces.

\no
Assuming that $R$ admits an expansion (``semiclassical'') in positive power series of a parameter (usually denoted
$\hbar$) as
\be
R_{12} = 1 \otimes 1 + \hbar r_{12} + {\cal O}(\hbar^2)\ ,
\label{classlimR}
\ee
the first non-trivial term arising when we substitute this in (\ref{YBR}) is of order two
and yields the classical YB equation
\be
[ r_{12}, r_{13}] + [r_{12}, r_{23}] + [r_{13}, r_{23}] = 0\ .
 \label{YBClR}
\ee
This is the canonically known ``classical Yang--Baxter equation''. It is not in general
the sufficient associativity condition for a classical linear Poisson bracket, except when $r$ is
non-dynamical and skew-symmetric (see e.g. \cite{STS}).
We shall hereafter limit ourselves to such situations.

\no
A quantum monodromy matrix $T$ is generically built as a tensor product over ``quantum spaces'' and algebraic
product over ``auxiliary space'' of representations of the YB algebra associated to $R$.
Namely, one assumes a collection operators assembled in matrices $L_{1i}$, acting on ``quantum''
Hilbert spaces labeled by $i$ and encapsulated
in a matrix ``acting'' on the auxiliary space $V_1$. For any quantum space $q$ they obey
the quadratic exchange algebra \cite{FTS, FT, tak}
\be
R_{12}\ L_{1q}\ L_{2q} =L_{2q}\ L_{1q}\ R_{12} \ ,
\label{YBRgen}
\ee
where operators acting on different quantum spaces commute.
The form of the monodromy matrix $T$ is then deduced from the co-module structure of the YB algebra
\be
T_a \equiv L_{a1}\ L_{a2}\ \ldots\ L_{aN}\
 \label{TM1}
\ee
and thus naturally obeys the same quadratic exchange algebra (\ref{YBRgen}).
In particular one can pick $L = R$,
the operators now acting on the second auxiliary space identified as ``quantum space''.
This way, one builds closed inhomogeneous
spin chains with general spins at each lattice site (labeled by $(i)$)
belonging to locally chosen representations of some Lie algebra (labeled by $i$).

\no
We now establish that $T$ has a classical limit by considering in addition the classical counterpart of
$L$, labeled by $L^c$ which then satisfies the quadratic Poisson algebra,
emerging directly as a semi-classical limit of (\ref{YBRgen}), after setting
$~{1\over \hbar} [A,\ B] \to \{ A,\ B \}$. It reads
\be
\{ L^c_a(\lambda_1),\ L^c_b(\lambda_2) \} =
[r_{ab}(\lambda_1 -\lambda_2),\ L^c_a(\lambda_1)\ L^c_b(\lambda_2)]\ .
\label{semicl0}
\ee
The quantum monodromy matrix has also a classical limit given by (see also \cite{ftbook, sklyaninlect})
\be
T^c_{a, \{i\}} =  L^c_{a 1}\ \ldots\ L^c_{a N}\ .
\label{classlimT}
\ee
The exchange algebra for $T^c$ takes the form
\be
\{T^c_a, T^c_b\} = [r_{ab},\ T^c_a\ T^c_b]\  .
\label{semicl}
\ee
This quadratic Poisson structure implies that the traces of powers of the monodromy
matrix $tr (T^c)$ generate Poisson-commuting quantities identified as classically integrable
Hamiltonians. Performing the trace over the finite vector space yields a generating
function ${\rm tr}(T^c(\lambda))$ for classically integrable Hamiltonians obtained by series expansion
in $\lambda$.

\subsection{The long wavelength limit}

The usual presentation of the long wavelength limit,
such as that found in \cite{fradkin, string}, is a Lagrangian one where the Poisson
structure is obtained from the standard derivation of canonical variables using a Lagrangian density.
Instead, we will present here a purely Hamiltonian version of this limit
by defining the long wavelength limit of a hierarchy of integrable quantum Hamiltonians
based on some affine Lie algebra $\hat{G}$.
We shall {\cal define a priori} the Poisson structure of the classical variables by imposing
classical integrability of the long wavelength limit of the Hamiltonian through its associated
classical Lax matrix.
We consider a $N$-site closed spin chain Hamiltonian $H$, initially
assumed to be governed by a nearest-neighbour interaction that takes the form
\be
H \equiv \sum_{1=1}^N H_{l l+1}\ .
\ee
The classical, long wavelength limit, is obtained  by first defining
local quantum states as linear combinations of the base
quantum states. The bras and kets are denoted respectively by
$\langle  n(l, \theta_k)|$ and $|  n(l, \theta_k) \rangle$, where $l$
denotes the site index and the $\theta_k$'s denote the
set of $k$ angular variables. The condition of ``closed'' spin chain, essentially formulated
as $N+l \equiv l$,  imposes
periodicity or quasi-periodicity conditions on the $\theta_k$'s. Note that we have assumed that the
base quantum states differ only by the fact that they are defined in distinct sites, hence the
frequently used notation below $|n_l\rangle$, instead of $|n(l,\th_k)\rangle$, should not be confusing.

\no
If one considers nearest-neighbor local interactions
then one defines the classical, but still defined on the lattice, Hamiltonian as
\be
{\cal H} \equiv \sum_{1=1}^N {\cal H}_{l}(t)\ ,\qq
{\cal H}_l(x, t) =  \langle  n_l| \otimes \langle n_{l+1}|\ H_{l l+1}\
| n_l\rangle\otimes |n_{l+1}\rangle\ .
\label{lwlH1}
\ee
For integrable models, we may similarly define the continuum limit of the full set of
commuting Hamiltonians. In these cases the generic Hamiltonians $H^{(n)}$
of the integrable hierarchy are obtained
{\cal directly} from the analytic
series expansion around some value $\lambda_{0}$ of the spectral parameter
of the trace of the monodromy matrix (transfer matrix) as
\be
{\rm tr} T(\lambda) \equiv \sum_{n=1}^{\infty} (\lambda -\lambda_{0})^n H^{(n)}\ .
\label{fhhjh1}
\ee
By extension, we define in this case the classical Hamiltonians as the expectation value, over the $N$ site
lattice quantum state, of $H^{(n)}$
\be
{\cal H}^{(n)}(x, t) =  \otimes_1^N \dots \langle  n_l| \otimes \langle n_{l+1}|\dots\ H^{(n)}\
\dots | n_l\rangle \otimes |n_{l+1}\rangle \dots \ .
\label{lwlH2}
\ee

\no
We next define a continuous limit and take simultaneously the thermodynamical limit in which $N \rightarrow \infty$.
Accordingly, this is achieved by identifying the lattice spacing
$\delta$ as being of order $1/N$ % (${\cal O}(\delta) \sim {\cal O}({1\over N})$).
and subsequently consider only slow-varying spin configurations (the long wavelength limit proper) for
which
\be
l_i \to l(x)\ ,\qq l_{i+1}  \to l(x +\delta)\ .
\label{cont1}
\ee
In this limit, the finite ``site differences'' turn into derivatives.

\no
Given that (\ref{lwlH2}) is applied to Hamiltonians of the integrable hierarchy obtained
{\cal directly} from the series expansion of the trace of the monodromy matrix, it is
immediate that the expectation value procedure goes straightforwardly to
the full monodromy matrix $T$ (and thence to its trace over the auxiliary space which is altogether
decoupled from the quantum expectation value procedure). Accordingly, we define first a
lattice expectation value
\be
T_a = \dots \langle  n_l| \otimes \langle n_{l+1}|...\ (L_{a 1}\ L_{a 2} \ldots L_{a N})
\dots | n_l\rangle \otimes |n_{l+1}\rangle \dots \ ,
 \label{lwlL2}
\ee
which nicely factors out as
\be
T_a =  \prod_{i=1}^N \langle  n_i|  L_{ai} |n_i\rangle \ .
\label{lwlLfact2}
\ee
Assuming now that $L$ admits an expansion in powers of $\delta$ as
\be
L_{ai}= 1 + \delta l_{ai} + {\cal O}(\delta^2)\ ,
\ee
we consider the product (setting $\langle n_i|l_{ai}|n_i \rangle = l_a(x_i)$)
\be
T_a = \prod_{i=1}^N (1 + \delta l_{ai} + \sum_{n=2}^{\infty} \delta^n l^{(n)}_{ai})\ .
\ee
Expanding this expression in powers of $\d$, we get
\be
T_a = 1 + \delta \sum_i l_{ai} + \delta^2\sum_{i<j} l_{a_{i}} l_{aj} + \delta^2 \sum_{i} l^{(2)}_{ai}
+\dots \ .
\ee
These, multiple in general, infinite series of the products of local terms,
are characterized by two indices: the overall power $n$ of $\delta$, and the number $m$
of the set of indices $i$ (that is the number of distinct summation indices)
over which the series is summed. Note that, in the $T$ expansion one always has $n\geqslant m$.
The continuum limit soon to be defined more precisely, will entail the limit $\delta \to 0$ with
${\cal O}(N) = {\cal O}({1/\delta})$.
We now formulate the following {\it power-counting rule}, that is terms of the form
(for notational convenience $l_{ai}=l^{(1)}_{ai}$ below)
\be
\delta^n \sum_{i_1< i_2 <\dots i_m}l_{ai_1}^{(n_1)} ...l_{ai_m}^{(n_m)}\ ,
\qq \sum_{j=1}^m n_j =n\ ,
\ee
with $n>m$ are omitted in the continuum limit. For a sinlge summation, the latter is defined by
\be
\delta \sum_{i} l_{ai} \to \int_{0}^{A} dx\ l_a(x)\
\ee
and similarly for multiple summations. Here $A$ is the length of the continuous interval
defined as the limit of $N \delta$.
In other words, contributions to the continuum limit may only come from the terms with $n=m$
for which the power $\delta^n$ can be exactly matched by the ``scale'' factor $N^m$ of the $m$-multiple sum over
$m$ indices $i$. In particular, only terms of order one
in the $\delta$ expansion of local classical matrices $L_{ai} \equiv \langle  n_i|  L_{ai}
|n_i\rangle$ will contribute to the continuum limit.
Any other contribution acquires a scale factor $\delta^{n-m} \to 0$, when the continuum limit is taken.
This argument is of course valid term
by term in the double expansion.
Being only a weak limit argument, it always has to be checked for consistency.

\no
Let's remark that if $L$ is taken to be $R$, one naturally identifies $\delta$ with
the small parameter $\hbar$, thus identifying in some sense the classical and the continuum limits.
However, this is not required in general. It is clear to characterize separately
both notions in our discussion as
\be
{\rm classical\ limit:} && R= 1 +\hbar r\ ,
\nonumber\\
{\rm continuum\ limit:} && L = 1 + \delta l\ .
\ee
Recalling (\ref{cont1}),
the continuous limit
of $T$, hereafter denoted ${\cal T}$,
is then immediately identified
from (\ref{lwlLfact2}), as the path-ordered
exponential from $x=0$ to $x={\mathrm A}$
\be
{\cal T} = P \exp{\left (\int_{0}^{A} dx\  l(x) \right )}\ ,
\ee
where suitable (quasi) periodicity conditions on the
continuous variables $\theta_k(x)$ of the classical matrix $l(x)$,
acting on the auxiliary space $V \otimes C(\lambda)$, are assumed.
Of course the
definition of a continuous limit requires that the $L$-matrices are not
\cal{too} inhomogeneous (e.g. $L$-matrices at neighbor sites should not be too different).
This is in fact assured by the long wavelength limit assumption.

\subsection{The Lax matrix and $r$-matrix formulation}

The above identification of ${\cal T}$
also defines it as the monodromy matrix of the first order differential operator
$d/dx + l(x)$. In addition, it has been built so as to straightforwardly
generate the classical continuous limit
of the Hamiltonians in (\ref{lwlH2}) from the analytic expansion
\be
{\rm tr} ({\cal T}(\lambda)) \equiv \sum_{n=1}^{\infty} (\lambda - \lambda_{0})^n {\cal H}^{(n)} \ .
\label{fhhjh2}
\ee
We thus characterize $l(x)$ as a local Lax matrix yielding
the hierarchy of continuous Hamiltonians ${\cal H}^{(n)}$. In order for this statement to agree with the
key assumption of preservation of integrability we are now lead to
require a Poisson structure for $l$ (inducing one for the continuous dynamical variables $\theta_k(x)$)
compatible with the demand of classical integrability of the continuous Hamiltonians.
Indeed, such a structure is deduced from (\ref{semicl0}), as the ultra-local Poisson bracket
\be
\{ l_1(x, \lambda_1),\ l_2(y, \lambda_2)\}
= [r_{12}(\lambda_1 -\lambda_2),\ l_1(x, \lambda_1)+l_2(y, \lambda_2)] \delta(x-y)\ ,
\label{funda1}
\ee
where $r$ is the classical limit (\ref{classlimR}) of the $R$-matrix characterizing the exchange algebra
of the $L$-operators.
More specifically, recalling that $L_{ai} = 1 + \delta l_{ai} + {\cal O}(\delta^2)$, plugging it into (\ref{semicl0})
and assuming ultra-locality of Poisson brackets one gets
\be
\{l_{ai},\ l_{bj}\} = [r_{ab},\ l_{ai} + l_{bj}]{\delta_{ij} \over \delta}\ .
\ee
One then identifies,  in the continuum limit $\delta \to 0$, the factor
 ${\delta_{ij}/ \delta}$ with $\delta(x-y)$.
\no
We then obtain a hierarchy of classically integrable, mutually Poisson commuting Hamiltonians
from the explicit computation of the transfer matrix $t(\lambda)$ of the Lax operator $d/dx + l(x)$
as $H^{(n)} = {d^n \over d \lambda^n} t(\lambda)|_{\lambda = {\lambda_0}}$. Such Hamiltonians are
however generally highly non-local and not necessarily very relevant as physical models. We shall
thus extend our discussion to local Hamiltonians.

\subsection{Local spin chains}

Local spin chain Hamiltonians are more interesting, physically meaningful and
easier to manipulate. In particular, they are the most relevant objects
in connection with string theory and the AdS/CFT duality \cite{miha}. Their construction generically requires the
determination of a so-called ``regular value'' $\lambda_0$ of the spectral parameter
such that $L_{ai}(\lambda_0) \propto {\cal P}_{ai}$, where ${\cal P}$ is the
permutation operator. In this sense the expansion of
$L$ can be expressed up to an appropriate normalization factor as
\be
L(\lambda) = f(\lambda) (1 + \delta l + {\cal O}(\delta^2)) \ .
\ee
Of course only when the auxiliary space $a$ and quantum
space $i$ are isomorphic has this ``regular value'' any relevance.
One then defines the local Hamiltonians as (denoting as usual $t(\lambda) = {\rm tr}_{a} T_a(\lambda)$)
\be
H^{(n)} = {d^n \over d \lambda^n} \ln (t(\lambda))\big |_{\lambda = {\lambda_0}} \ ,
\label{locham}
\ee
Their long wavelength limit (e.g. (\ref{lwlH1}))
is not obviously derivable from a straightforward ``diagonal''
expectation value of the $T$-matrix contrary to (\ref{lwlH2}), since
in general $\langle F(A)\rangle \neq F(\langle A \rangle)$, for any functional of
a set of operators $A$. However, we show below that this is
indeed the case due to locality properties.
Let us first focus for simplicity (but, as we shall see, without loss of generality)
on the first local Hamiltonian
\be
H^{(1)} = t(\lambda_0)^{-1}{d \over d \lambda} t(\lambda)\big |_{\lambda = {\lambda_0}}\ ,
\label{locham2}
\ee
where, $t^{-1}(\lambda_0) = {\cal P}_{12}{\cal P}_{23}\dots {\cal P}_{N-1N}$. This operator
acts exactly as a one-site shift on tensorized states, identifying of course site labels
according to the assumed periodicity, i.e. $N+1 = 1$.
(Normalization issues are discussed in \cite{ads}).
Computing the expectation value of $H^{(1)}$ we obtain
\be
\langle  H^{(1)} \rangle = \langle n_1| \otimes \ldots \otimes \langle n_N| t^{-1}(\lambda_0)
{d\over d\lambda}\Big (f^N(\lambda) Tr_a\prod_{i=1}^N(1 + \delta l_{ai} + {\cal O}(\delta^2)) \Big)
|n_1\rangle \otimes \ldots \otimes |n_N\rangle \ .
\ee
One has
\be
\langle n_1| \otimes \langle n_2 | \otimes \ldots \langle n_N| t^{-1}(\lambda_0) =
\langle n_2| \otimes \langle n_3 | \otimes \ldots \langle n_1|
\ee
and of course $N+1 \equiv 1$.

\no
Taking into account the {\it power-counting rule} described in section 2.2
we obtain that
\be
\langle H^{(1)} \rangle
= \prod_{i=1}^N \langle n_{i+1}|n_i\rangle {d\over d\lambda}
\Big (f^N(\lambda){\rm tr}_a\prod_{i=1}^N (1 + \delta \langle\ l_{ai}\ \rangle+
{\cal O}(\delta^2)) \Big )\ .
\ee
We then easily establish that in the continuum limit,
using the power counting rule and the factorized form of both the state vector
as $\langle n_1| \otimes \ldots \otimes \langle n_N|$ that the operator to
be valued over it $t^{-1}(\lambda_0) = {\cal P}_{12}{\cal P}_{23}\dots {\cal P}_{N-1N}$,
$\langle t^{-1}(\lambda_0) \rangle= \langle t(\lambda_0)\rangle^{-1}$.
We finally obtain that in the continuum limit
\be
\langle H^{(1)}\rangle = \langle t^{-1}(\lambda_0) {d\over d \lambda} t(\lambda)\big \vert_{\lambda=\lambda_0} \rangle =
\langle t(\lambda_0)\rangle^{-1}{d\over d \lambda} \langle t(\lambda)\rangle \big \vert_{\lambda= \lambda_0} =
{d \over d \lambda}(\ln \langle t(\lambda)\rangle )\big \vert_{\lambda = \lambda_0}\ .
\ee
The computation may be easily generalized along the same lines for any higher Hamiltonian.
Higher local Hamilltonians are indeed obtained from \eqn{locham},
%\be
%H^{(n)} = {d^n\over d \lambda^n}\ln(t(\lambda))\big \vert_{\lambda =\lambda_0}\ ,
%\ee
admitting thus an expansion as
\be
H^{(n)} = t^{-1}(\lambda_0) {d^n\over d \lambda^n}t(\lambda)\big \vert_{\lambda_0}+ \mbox{polynomials}\ ,
\ee
depending only on lower order local Hamiltonians.
When computing the expectation value of such higher Hamiltonians one gets
the expectation value of $t^{-1}(\lambda_0) {d^n\over d \lambda^n}t(\lambda)\vert_{\lambda_0}$ which in
the continuum classical limit yields
\be
\langle t^{-1}(\lambda_0) {d^n\over d \lambda^n}t(\lambda)\vert_{\lambda_0}\rangle =
\langle t(\lambda_0) \rangle^{-1} {d^n\over d \lambda^n}\langle t(\lambda) \rangle \vert_{\lambda_0}\ ,
\ee
using the same arguments as in the $n=1$ case.
In addition, one obtains expectation values of the polynomials of order $k$ in the local Hamiltonians.
In this case expectation values by tensor product of local
vectors $\langle n_1| \ldots \langle n_N|$ are exactly factorized over products of $k$
local monomials $h_{i_1}\dots h_{k_k}$,
except if some of the indices $i$ coincide (or at least overlap for multiple indices).
Locality of the lower Hamiltonians plays here a crucial role.
It is clear that such families of terms with coinciding or overlapping
indices correspond to a second ``label''
$M = k-1$ and therefore their contribution will necessarily be suppressed in the continuum limit,
with respect to the contribution of the
generic terms (non-coinciding indices) with $M =k$ by the power-counting argument.
Hence, it is consistent to conclude that in the continuum limit
\be
\langle \mbox{Polynomial\ in}\ (H^{(i)}) \rangle =  \mbox{Polynomial\ in}\ (\langle H^{(i)} \rangle)
\ee
and therefore
\be
\boxed{
\
\langle H^{(n)}\rangle  =\langle
{d^n\over d \lambda^n}\ln(t(\lambda)) \Big |_{\l=\l_0}\rangle
= {d^n\over d \lambda^n}\ln(\langle t(\lambda)\rangle) \Big |_{\l=\l_0}
\ }
\ .
\label{loccharges}
\ee
This is the final, key result in systematically establishing the classical continuum limit of integrable spin chains.
We may now apply this general procedure to all sorts of examples, starting with the
simpler applications.

\section{Examples}

\subsection{The isotropic Heisenberg model}

The isotropic Heisenberg model (XXX chain) Hamiltonian describing first neighbor
spin-spin interactions is given by
\be
H = {1\over 2} \sum_{j=1}^N \Big ( \sigma^x_j \sigma^x_{j+1} + \sigma^y_j
\sigma^y_{j+1} +  \sigma^z_j \sigma^z_{j+1} \Big )\ .
\ee
It is well known that when one considers the long wavelength limit
one obtains a classical $\s$-model \cite{fradkin, string}. We shall briefly review how this process works.
The coherent spin state is parametrized by the parameters $x,\ t$ via the fields $\theta, \ \varphi$ as
\be
|n(x, t)\rangle = \cos  \theta(x, t)\
e^{i \varphi(x, t) }\ | + \rangle \ +\  \sin  \theta(x, t)\ e^{-i \varphi(x, t) }\ | - \rangle\ ,
\ee
where the ranges of variables is $\th\in (0,\pi/2)$ and $\varphi\in (0,\pi)$.
One can verify the completeness relation
\be
\int d\mu( n) | n\rangle \langle n| = 1\ ,
\ee
where the integration measure is given by
\be
d \mu (n) = {4 \over \pi}\ \sin\theta \ \cos\theta\ d \theta\ d \varphi\ .
\ee
Then as was described in \cite{fradkin, string} and in subsection 2.1, one obtains a classical Hamiltonian
via the expectation value procedure by employing \eqn{lwlH1}.
%\be
%{\cal H} =  \langle  n_l| \otimes \langle n_{l+1}|\ H_{l l+1}\
%| n_l\rangle\otimes |n_{l+1}\rangle\ .
%\ee
The appropriate XXX 2-site Hamiltonian is
\be
H_{l l+1} \propto ({\cal P}_{l l+1} -{\mathbb I})\ ,
\label{h1}
\ee
where $\cal P$ is the permutation operator acting as
${\cal P} (a\otimes b) = b \otimes a$ for $a,\ b$ vectors in $V$.
From the definition of ${\cal H}$ we are led to compute quantities of the type
\be
\langle a| \otimes \langle b|\ {\cal P}\ |a \rangle \otimes |b\rangle = \langle a|b \rangle
\otimes \langle b |a\rangle = |\langle a|b \rangle|^2\ .
\ee
They are expressed in terms of scalar products of the form
\be
\langle  \tilde n|  n \rangle = \cos(\theta - \tilde \theta)\ \cos(\varphi - \tilde \varphi) +
i \cos(\theta + \tilde \theta)\ \sin(\varphi - \tilde \varphi)\ .
\ee
In the long wavelength limit, $| n \rangle - |\tilde n \rangle = |\delta n \rangle $,
$\tilde \theta(x)= \theta(x +\delta)$ and  $ \tilde \varphi(x) = \varphi(x + \delta)$. We finally conclude that
\be
H \propto \int d x \
(\theta^{'2}+\sin^2(2 \theta)\ \varphi^{'2})\ .
\label{H0}
\ee
We shall now derive the Lax representation yielding (\ref{H0}) following section 2.
The $R$-matrix for the XXX  model is \cite{yang}
\be
R(\lambda) = \lambda + i\hbar {\cal P}\ .
\label{r1}
\ee
This $R$-matrix is a solution of the
quantum YB equation \cite{baxter}.
It has a consistent normalized classical limit defined as
\be
r(\lambda) = {1 \over \lambda} {\cal P}\ ,
\ee
which satisfies the classical YB equation. Alternatively,
the classical $r$-matrix may be written as
\be
r(\lambda) =  {1 \over  \lambda} \begin{pmatrix}
    {1\over 2}(\sigma^z+1) & \sigma^-  \\
    \sigma^+ & {1\over 2}(-\sigma^z+1)
\end{pmatrix} \  .
\ee
Set first
\be
L_{an}(\lambda) =R_{an}(\lambda -{i\hbar \over 2})\
\label{lan}
\ee
and demand that $L$ satisfies the fundamental algebraic relation
\be
R_{ab}(\lambda_1 -\lambda_2)\ L_{an}(\lambda_1)\ L_{bn}(\lambda_2)=
L_{bn}(\lambda_2)\ L_{an}(\lambda_1)\ R_{ab}(\lambda_1 -\lambda_2)\ ,
\label{funda0}
\ee
where as usual in the spin chain framework we call $n$ the quantum space and
$a$ the auxiliary space.
Following the general derivation of section 2 and going directly to
the continuous limit we disregard higher powers in $\delta = \hbar$ (in this case the two
small parameters are naturally identified).
We next define a ``local Lax matrix" as a mean value of $L$
on the same coherent spin state, taken solely over the quantum space
\be
\langle n| L_{an}(\lambda) | n \rangle  =   1 +i \hbar l(x,\lambda) \ ,
\ee
where
\be
l =  \begin{pmatrix}
   {1\over 2}\langle n|\sigma^z |n \rangle  & \langle n| \sigma^-  |n \rangle \\
    \langle n|\sigma^+ |n \rangle & -{1\over 2}\langle n|\sigma^z |n \rangle
\end{pmatrix}
= \ha
\begin{pmatrix}
   \cos 2 \theta (x)  & \sin 2 \theta(x)\ e^{- 2i \varphi(x)} \\
  \sin 2 \theta(x)\ e^{+ 2i \varphi(x)}& -\cos 2 \theta (x)
\end{pmatrix}\ ,
\label{llxxx}
\ee
where we have used the form of the coherent states to compute the matrix elements explicitly.
%\be
%\langle n|\sigma^z |n \rangle =  \cos 2 \theta (x)\ ,\qq
%\langle n|\sigma^{\pm} |n \rangle = {1\over 2} \sin 2 \theta(x)\ e^{\mp 2i \varphi(x)}\ .
%\ee
Then $l$ satisfies the
classical fundamental algebraic relation
\be
\{l_1(x, \lambda_1),\ l_2(y, \lambda_2)\} = [r_{12}(\lambda_1 -\lambda_2),\ l_1(\lambda_1)+l_2(\lambda_2)]\delta(x-y)\ .
\label{funda}
\ee
Setting $l(x, \lambda)=  \Pi/\l$
and taking into account the above algebraic relations we get
\be
\{\Pi_1,\ \Pi_2\} =  {\cal P}_{12} (\Pi_2 -\Pi_1)\delta(x-y)\ .
\ee
The parametrization in terms of the continuum parameters $\theta(x)$, $\phi(x)$
gives rise to the classical version of $sl_2$. Indeed,
parametrizing the generators of the classical current algebra as
\be
S^z =  \cos 2 \theta\ , \qq S^{\pm}= {1\over 2} \sin 2 \theta\ e^{\mp 2i \varphi}\ .
\label{par1}
\ee
we obtain from the fundamental relation that
\be
\{S^{+},S^{-}\} =  S^z\delta(x-y)\ , \qq \{S^z, S^{\pm} \} = \pm 2 S^{\pm}\delta(x-y)\ .
\label{poisson1}
\ee
The continuum parameters $\theta(x)$ and $\phi(x)$ can also be expressed in terms of canonical variables $p$ and
$q$ as
\be
\cos 2 \theta(x) = p(x)\ , \qq
\varphi(x) = q(x) ~~~\mbox{and} ~~~~\{q(x),\ p(y)\} = i \delta(x-y)\ .
\label{par2}
\ee
The $l$-matrix in \eqn{llxxx}
coincides obviously with the potential term in the Lax matrix of the classical Heisenberg model.
Precisely, one recalls that one must consider as classical Lax operator a la Zakharov--Shabat $L = d/dx + l(x)$.
The monodromy matrix for $L$ is well known now to yield the classical Hamiltonians including the first non trivial one
(see \cite{ftbook})
\be
\boxed{\
H \propto \int d x\ \left( \left({d S^z \over dx} \right)^2 +  \left({d S^x \over dx} \right)^2
+ \left({d S^y \over dx} \right)^2\right)
\ }\ .
\ee
Recalling (\ref{par1}) and substituting in the expression above we obtain the Hamiltonian (\ref{H0}),
hence the process above works consistently.

\no
Having exemplified the general construction of Section 2 to a simple system and checked the consistency
of the approach we now turn to more complicated
systems by first moving to trigonometric and elliptic $sl(2)$ $R$-matrices,
corresponding to the XXZ and XYZ spin chains.

\subsection{The anisotropic Heisenberg model}

Consider the generic anisotropic XYZ model with Hamiltonian
\be
H = {1\over 2} \sum_{j=1}^N \Big ( J_x \sigma^x_j \sigma^x_{j+1} + J_y\sigma^y_j
\sigma^y_{j+1} +  J_z\sigma^z_j \sigma^z_{j+1} \Big )\ .
\ee
For the following computations it is convenient to set
\be
J_{\xi} = 1 - \delta^2 a_{\xi}\ , \qq \xi \in \{x,\ y,\ z \}\ .
\ee
The Hamiltonian is written as
\be
H=  \sum_{j=1}^N{\cal P}_{j j+1} -{N\over 2} - {\delta^2 a_x \over 2} \sum_{j=1}^N \sigma^x_j \sigma^x_{j+1}
- {\delta^2 a_y \over 2} \sum_{j=1}^N \sigma^y_j \sigma^y_{j+1}
- {\delta^2 a_z \over 2} \sum_{j=1}^N \sigma^z_j \sigma^z_{j+1} \ .
\label{hdef}
\ee
The additive constant may be omitted here. Taking into account equations (\ref{h1})--(\ref{H0}), (\ref{hdef})
and keeping terms of order $\delta^2$ we get
\be
H \propto \int dx\  \Big ( \theta^{'2}+\sin^2(2 \theta)\ \varphi^{'2} +
a_x\sin^2 (2 \theta) \cos^2 (2\varphi)+
a_y\sin^2 (2 \theta) \sin^2 (2\varphi)+ a_z \cos^2( 2 \theta) \Big )\ .
\ee
This may be seen as an anisotropic ``deformation'' of the classical Heisenberg Hamiltonian.
The last three terms are essentially
potential-like terms. In the special case of the XXZ model the terms with coupling constant
$a_x, a_y$ are zero, whereas in the XXX case
all potential terms vanish and one recovers the Hamiltonian (\ref{H0}).
If we now recall the parametrization (\ref{par1}),
then the expression above reduces
to the Hamiltonian of the Landau-Lifshitz model or the anisotropic classical magnet \cite{ftbook}
\be
\boxed{\
H \propto \int dx\ \left ( \left({d S_z \over dx} \right)^2 +  \left({d S_x \over dx} \right)^2 +
\left({d S_y \over dx} \right)^2 +  a_xS_x^2 +a_y
S_y^2 + a_z S_z^2\right )
\ }
\ .
\ee
We now derive the classical $l$-matrix
for the anisotropic cases. We focus in more detail on the XXZ $R$-matrix
\be
R(\lambda) = \begin{pmatrix}
    \sinh (\lambda +{i \mu \over 2}\sigma^z+{i\mu \over 2}) & \sinh (i\mu)\sigma^-  \\
    \sinh (i\mu)\sigma^+ & \sinh (\lambda -{i \mu \over 2}\sigma^z+{i\mu \over 2})
\end{pmatrix}\ .
\ee
The classical limit of the XXZ $R$-matrix, after appropriate normalization, is given as
(we divide with the constant factor $\sinh \l$)
 \be
R(\lambda) = 1 + i \mu\ r(\lambda)+{\cal O}(\mu^2)\ ,
\ee
where
\be
r(\lambda) = {1\over \sinh\lambda} \begin{pmatrix}
     ({\sigma^z \over 2} +{1 \over 2})\ \cosh \lambda & \sigma^-  \\
    \sigma^+ &  (-{\sigma^z \over 2} +{1 \over 2})\ \cosh \lambda
\end{pmatrix}\ .
\ee
The associated classical Lax operator is again obtained from $L(\lambda) = R(\lambda - {i\mu \over 2})$ as
(once again moving immediately to the continuous limit)
\be
\langle n| L(\lambda)|n \rangle  =1+ i \mu\ l(x,\lambda)+ {\cal O}(\mu^2)\ ,
\ee
with
\be
l(\lambda) &=&  {1 \over \sinh\lambda} \begin{pmatrix}
\langle n| {\sigma^z \over 2}|n\rangle \ \cosh \lambda & \langle n|\sigma^-|n\rangle  \\
\langle n|\sigma^+|n\rangle &  -\langle n|{\sigma^z \over 2}|n\rangle \ \cosh \lambda
\end{pmatrix} \non\\&=& {1\over \sinh\lambda} \begin{pmatrix}
{1\over 2}S^z \cosh \lambda & S^-  \\
S^+ &  -{1\over 2}S^z\ \cosh \lambda
\end{pmatrix}\ ,
\ee
where $S^Z,\ S^{\pm}$ are the classical generators of the current $sl(2)$ algebra realized in terms of
the angular variables in \eqn{par1}.
The continuous variables $x,y$ were omitted here for simplicity and will be from
now on whenever there is no ambiguity.

\no
Let us also briefly characterize the classical algebra underlying the model.
We set
\be
l_i(\lambda) = {\cosh \lambda \over \sinh \lambda} D_i + {1\over \sinh(\lambda)} A_i\ ,\qq
r_{12}(\lambda)
= {\cosh \lambda \over \sinh \lambda} {\cal D}_{12} + {1\over \sinh(\lambda)} {\cal A}_{12}\ .
\ee
Substituting this expressions to (\ref{funda}) and taking into account that
\be
[{\cal A}_{12},\ A_1] = - [{\cal D}_{12},\ A_2]\ ,
\ee
we end up with the following set of Poisson structures
\be
\{D_1,\ D_2\} =0\ , \quad
\{D_1,\ A_2\}= [{\cal D}_{12},\ A_2]\delta(x-y)
\ , \quad \{A_1,\ A_2\}= -[{\cal A}_{12},\ D_1]\delta(x-y)\ ,
\ee
which give rise to the $sl_2$ Poisson algebra (\ref{poisson1}).

\no
The full XYZ classical $r$-matrix also yields, through
this process, the classical Lax operator of the fully anisotropic classical Heisenberg model,
satisfying also the fundamental linear algebraic relation (\ref{funda}) (see also \cite{ftbook}).
A detailed presentation of this derivation is omitted here for the sake of brevity.

\no
Generalizations of the Heisenberg model associated to higher rank algebras are also
presented in \cite{ads}, where a more detailed analysis of the described process can be found.

\end{document}